# ELECTRONIC MODEL OF A VOLCANO-SHAPED SIZE DEPENDENCE IN CO OXIDATION BY GOLD NANOPARTICLES

*Nigora Turaeva, Herman Krueger

Department of Biological Sciences, Webster University

470 E. Lockwood Ave., Webster Groves, MO 63119, USA

**Abstract**

The theoretical model of a volcano-type size dependence of extraordinary catalytic activity of gold nanoparticles in CO oxidation is proposed based on the electronic theory of catalysis, the jellium model of metal clusters and the d-band model of formation bonds in transition metals. The kinetics of the general mechanism of CO oxidation reaction catalyzed by gold nanoparticles through dissociative adsorption of oxygen molecules and interaction with CO molecules from the gas-phase is considered by introducing electronic steps into 3-step mechanism of the reaction, corresponding to transitions between weak and strong chemisorbed states of intermediates. By introducing the exponential dependences of the fractions of weak and strong chemisorbed intermediates (O, $CO_2$) on the size dependent Fermi level of the metal cluster with respect to antibonding energy levels of the intermediates, a volcano-type dependence of reaction rate on the size of gold nanoparticles is obtained. The dependence of the reaction rate upon the supports is taken into account by introducing the additional energy term into the Fermi level of the metal catalyst.

1. **Introduction**

One of the unique possibilities enabled by nanotechnologies is the control of properties of nanoparticles by their size [1], as opposed to approaches where materials are replaced by new materials or modified by impurities to achieve desirable properties. It is remarkable that at the nanometer scale materials can exhibit unique properties completely different from their bulk counterparts. One such material is chemically inert bulk gold, becoming active catalysts at the nanometer scale with respect to many oxidation and hydrogenation chemical reactions at milder conditions than traditional catalysts [2-18]. CO oxidation is one of the important oxidation reactions catalyzed by gold NPs, where non-monotonic size effects were experimentally observed in the range of 1-5 nm [3-12]. Other factors established experimentally that contribute to the reaction activity of gold NPs include the nature of the support [5,6,10,11], low coordinated gold atoms [8-10], defects presented at the interfaces [14,15], a non-metallic nature of gold clusters [5] and photon-induced or injected hot electrons [17,18]. In spite of intense research in this field, the origin of the size effects has not been completely established yet although many important aspects of the reaction were theoretically investigated [8-10, 14,15,19,20]. In our previous works [19,20], we obtained the agreement of the electronic model of the size dependence of CO oxidation reaction rate based on the jellium model of metal clusters and the Wolkenstein`s approach on the role of the Fermi level in catalytic reactions originally proposed for semiconductors within the mechanism of the reaction when oxygen molecules are adsorbed via dissociation into atoms. However, the chemical bonds formed between the adsorbates and the surface of the metal catalyst were not taken into account in the theory. In this study, we will show that a volcano-type size dependence of catalytic activity of gold NPs can be theoretically obtained based on the Wolkenstein`s approach but taking into account the d-band model explaining the origination of formation of chemical bonds between the adsorbed molecules and the transition metal catalyst. By taking into account the size dependence of the ionization potential of metal clusters and introducing weak and strong chemisorpsion states of intermediates of the reaction in the kinetic equations, a volcano-type size dependence of the reaction rate is obtained.

* nigoraturaeva82@webster.edu

## 2. Wolkenstein`s approach

According to the electronic theory of catalysis in semiconductors developed by Wolkenstein Th. [21,22], the chemisorbed particles are considered as structural defects of the crystal surface, introducing electron energy levels into the bandgap of semiconductors. Such structural defects of the surface can serve as localization centers for free electrons and holes. The localization of an electron or a hole on the chemisorbed particle causes a change of the character of its chemical bond with the surface. Thus, different forms of chemisorption of intermediates, from weak chemisorption to strong-acceptor and strong-donor chemisorption, are introduced. The weak chemisorption of particles takes place without participation of free valences of the surface (electrons or holes). In strong chemisorption, the free valences of the catalysts are involved to form chemical bonds with molecular orbitals of the adsorbed molecules. Electrons of the catalyst are responsible for forming acceptor bonds with the adsorbed molecules while positive charges of the surface form donor bonds. The intermediates exhibit different reactivity with respect to a specific reaction, and their relative numbers on the surface depend upon the Fermi level of the catalyst. In general, according to the electronic theory, a number of characteristics of chemical reactions catalyzed by semiconductors depend upon the Fermi level of the catalyst: (1) the relative number of different forms of chemisorption at electronic equilibrium, hence the reactivity; (2) the adsorptivity of the surface for each species of particles and the catalytic activity of the surface with respect to a given reaction; (3) the selectivity of a catalyst with respect to two (or more) concurrently proceeding reactions. In chemistry, the Fermi level is more familiar as electrochemical potential for electrons, the thermodynamic work required to add one electron to the system. In solid state physics, the Fermi level is considered to be as an energy level of an electron having a ½ probability of being occupied at thermodynamic equilibrium and responsible for electrical properties of the solids. In metals, the Fermi level is the boundary between occupied and unoccupied states, determining its work function.

The results of the electron theory of catalysis can fully be applicable to dielectrics, but it cannot be automatically applied to metallic catalysts, especially for chemical processes on transition metal surfaces. In this work we will apply the electronic theory to metallic catalysts by using the d-band model [23,24] explaining the origination of formation of covalent bonds between the adsorbed molecules and the transition metal catalyst. By comparison of the Fermi level of the metal and the antibonding molecular orbitals of the adsorbed molecules, the fractions of reactive intermediates can be evaluated which will determine the reaction rate of the catalytic reaction.

### 3. The d-band model

The d-band model [23,24] explains the origin of the chemical bond between the surface of transition metal atoms and the molecule adsorbed onto this surface. According to the model, the adsorbate interacts with the transition metal surface, resulting in two different effects, including a shift and broadening of the adsorbate valence states due the coupling with the s states of the metal, and formation of bonding and antibonding states with narrow distributions due to the coupling with the sharp and intense d-states of the metal. The adsorption is determined by the Fermi level of the metal relative to the energy of the antibonding states of the adsorbate. If the antibonding state is filled by electrons transferred from the aforementioned Fermi level, then the adsorption of the molecule does not occur. If the Fermi level is below the antibonding state, then it is empty. The molecule is adsorbed and forms a strong chemical bond with the surface. The lower the Fermi level lies with respect to the antibonding state, the stronger the bond.

It is well known [25,26] that the ability of metal clusters to accept or donate charge can be tuned by varying their size. According to the spherical jellium theory of metal clusters [23-25], the ionization potential and the electron affinity depend upon the size of clusters according to the formulas:

$$I(R) = W_b + \alpha \frac{e^2}{4\pi\varepsilon_0 r} + O(r^{-2}) \qquad (1)$$

$$A(R) = W_b - \beta \frac{e^2}{4\pi\varepsilon_0 r} + O(r^{-2}) \qquad (2)$$

Here $W_b$ is the work function of bulk metal, $r = r_s N^{1/3}$ is the radius of the cluster, $r_s$ is the Wigner-Seitz radius, $N$ is the number of atoms in the cluster, $\alpha$, $\beta$ are the parameters, which are equal to 3/8 and 5/8, respectively. Quantum effects implied in terms of higher orders (quadratic and higher) must be included for small clusters, $N<100$ (r<1.5nm) [25]. Since the Fermi level of gold varies between its ionization potential of 9.2 eV for a single atom and the work function of the bulk, 5.3 eV [1,27], its chemical properties can be tuned over a very wide range by variation of the cluster size.

Based on the d-band model, we can suppose that adsorption and desorption of molecules can be tuned by the size of metal nanoparticles. Here we should note that the turnover frequency of catalyzed chemical reactions as a function of binding energy of the reactants onto the catalyst surface yields a volcano-shaped plot according to Sabatier principle and Norskov et.al. [1,28]. The effect supposes that at low binding energies the reactant is weakly absorbed to form the transition state of the reaction while at high adsorption energies the product does not desorb. Extending these ideas to the CO oxidation by gold NPs, on the basis of the d-band model, we suppose that by decreasing the size of nanoparticles the Fermi level moves down below the antibonding states of the reagents on the surface and the adsorption of $O_2$ molecules are enabled. The number of adsorbed molecules increases upon decreasing the particle size. We suppose that desorption of the products continues to take place until the size-dependent Fermi level goes below the antibonding states of the product ($CO_2$ molecules). Further decrease in particle size leads to the impairment of desorption of the products. So, there is an optimal size range of nanoparticles, when a compromise between adsorption of reagents and desorption of products is reached. For gold nanoparticles, it is known from the experiments that the optimal size range is between 1-5 nm. Note also that for bulk gold it is known that adsorption of $CO_2$ molecules is mostly of a physical character [29].

By means of kinetic equations for reaction intermediates, the aforementioned size effects can be quantified. However, the mechanism of the reaction is unclear yet as to whether the reaction occurs through the adsorption of $O_2$ or CO molecules or through the absorption of both reagents. In our previous study [19], the expression for the non-monotonic size dependence of the catalytic activity of gold NPs with respect to CO oxidation was obtained in the framework of kinetic equations for the mechanism of the reaction, when the adsorption of oxygen molecules and the reaction with the CO molecules from the air. In the model, the jellium model of metal clusters was taken into account but without consideration of the concepts of the d-band model on the formation of chemical bonds between the adsorbate and the surface. In this study, we derive the expression for the volcano-type size dependence of catalytic activity of gold NPs within the mechanism of the reaction through dissociative adsorption of $O_2$ molecules which becomes energetically favorable at the nanometer range [10] by taking into consideration the d-band model.

## 4. The size dependence of the reaction rate

We will consider a three-step mechanism of CO oxidation reaction (Table 1): (1) dissociative adsorption of $O_2$ molecules onto the surface of the catalyst; (2) interaction of atomic oxygen with CO molecules from the gas phase and formation of $CO_2$ molecules; (3) desorption of $CO_2$ molecules. Into this scheme, we will include two electronic steps (indicated with asterisk), corresponding to transitions between weak (neutral) and strong (negatively charged) chemisorption states of intermediates. The oxygen adsorbates are activated by electron transfer from the catalyst. Being in the ion-radical state, the negatively charged O atoms react with CO molecules to form ion-radicals $CO_2$ molecules as intermediate compounds. $CO_2$ molecules being neutralized desorb from the surface of the catalyst. We will not include reverse reactions of gas-phase adsorption of products.

| | |
|---|---|
| $O_2 + 2Z \leftrightarrow 2ZO^0$ | (1) |
| $ZO^0 \leftrightarrow ZO^-$ | (1*) |
| $ZO^- + CO \leftrightarrow ZCO_2^-$ | (2) |
| $ZCO_2^- \leftrightarrow ZCO_2^0$ | (2*) |
| $ZCO_2^0 \rightarrow CO_2 + Z$ | (3) |

Table 1. The general mechanism of CO oxidation with two electronic steps

Here Z is the active site of the catalyst. This general mechanism of the reaction corresponds to the following valence signs scheme of chemisorption forms of the intermediates (Fig. 2).

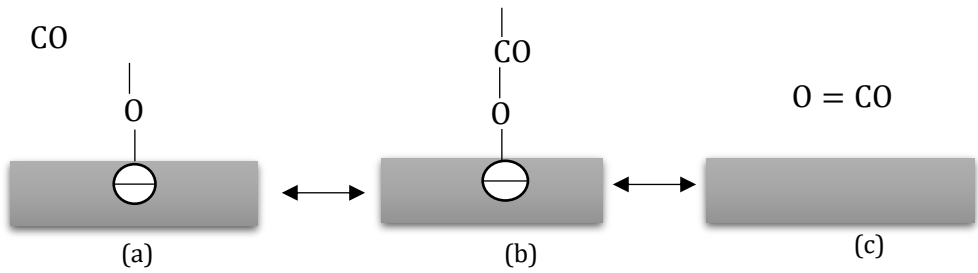

Fig. 2. The valence signs scheme of CO oxidation through adsorption of oxygen (a), formation of $CO_2$ (b) and desorption of $CO_2$ (c). The "minus" sign indicates the charged state of the intermediates, corresponding to the strong chemisorption.

Since electronic processes occur at much faster rate than atomic rebuildings, we assume that the electronic equilibrium is established, and the intermediate fractions characterize the equilibrium relative amounts of the different forms of chemisorption on the surface. Then, we can introduce the fractions of chemisorbed molecules in different charged states [21]:

$$\eta_O^- = \frac{N_O^-}{N_O^- + N_O^0}; \quad \eta_O^0 = \frac{N_O^0}{N_O^- + N_O^0}; \quad \eta_{CO2}^- = \frac{N_{CO2}^-}{N_{CO2}^- + N_{CO2}^0}; \quad \eta_{CO2}^0 = \frac{N_{CO2}^0}{N_{CO2}^- + N_{CO2}^0}, \quad (3)$$

where $N$ is the concentration of intermediates (O, $CO_2$) on the surface in neutral and negatively charged states, corresponding to weak and strong chemisorption, respectively. Then the kinetic equations for the intermediates of the reaction take the following forms:

$$\begin{cases} \dfrac{dN_O}{dt} = k_1 P_{O_2}(1 - s(N_O + N_{CO2}))^2 - 2k_{-1}(N_O^0)^2 - k_2 N_O^- P_{CO} + k_{-2} N_{CO2}^- & (4) \\ \dfrac{dN_{CO2}}{dt} = k_2 N_O^- P_{CO} - k_{-2} N_{CO2}^- - k_3 N_{CO2}^0 & (5) \end{cases}$$

Here P is the pressure of the reagents, s is the effective surface area of a single active site. At the quasi steady-state conditions, we have the following equations:

$$\begin{cases} k_1 P_{O_2}(1 - s(N_O + N_{CO2}))^2 - 2k_{-1}(N_O^0)^2 - k_3 N_{CO2}^0 = 0 & (6) \\ k_2 N_O^- P_{CO} - k_{-2} N_{CO2}^- - k_3 N_{CO2}^0 = 0 & (7) \end{cases}$$

By assuming $2k_{-1}(N_O^0)^2 \gg k_3 N_{CO2}^0$, we can find $N_O$ from Eq. (6) by taking into account the expressions (3)

$$N_O = \dfrac{1 - s N_{CO2}}{s + \eta_O^0 \sqrt{\dfrac{2k_{-1}}{k_1 P_{O_2}}}}. \tag{8}$$

Inserting Eq. (8) into Eq. (7), the following formula for the reaction rate can be obtained

$$R = k_3 N_{CO2}^0 = \dfrac{k_2 k_3 P_{CO}}{k_2 P_{CO} s \dfrac{1}{\eta_{CO2}^0} + \dfrac{1}{\eta_O^-}\left(s + \eta_O^0 \sqrt{\dfrac{2k_{-1}}{k_1 P_{O_2}}}\right)\left(k_{-2} \dfrac{\eta_{CO2}^-}{\eta_{CO2}^0} + k_3\right)}. \tag{9}$$

Using the Fermi statistics and the concepts of the d-band model discussed above, we can define those fractions as a function of antibonding states of adsorbed intermediates with respect to the Fermi level of gold nanoparticles $E_f$ by the following expressions

$$\eta_O^- = \dfrac{1}{1 + e^{-(E_f - E_{O-Au}^a)/k_B T}}; \tag{10}$$

$$\eta_O^0 = \dfrac{1}{1 + e^{(E_f - E_{O-Au}^a)/k_B T}}; \tag{11}$$

$$\eta_{CO2}^- = \dfrac{1}{1 + e^{(E_f - E_{CO2-Au}^a)/k_B T}}; \tag{12}$$

$$\eta_{CO2}^0 = \dfrac{1}{1 + e^{-(E_f - E_{CO2-Au}^a)/k_B T}}; \tag{13}$$

$$\dfrac{\eta_O^0}{\eta_O^-} = e^{-\dfrac{E_f - E_{O-Au}^a}{k_B T}}; \tag{14}$$

$$\dfrac{\eta_{CO2}^-}{\eta_{CO2}^0} = e^{(E_f - E_{CO2-Au}^a)/k_B T}. \tag{15}$$

Here $k_B$ is the Boltzmann`s constant, $T$ is the temperature, $E_{O-Au}^a$ and $E_{CO2-Au}^a$ are the energies of antibonding states of O and $CO_2$ on the surface of the catalyst, respectively. The size dependent Fermi level of the catalyst is determined by the ionization potential of the metal cluster from Eq. (1) by introducing the additional term $E_{sup}$, which takes into account the role of supports

$$E_f = W_b + \alpha \frac{e^2}{4\pi\varepsilon_0 r} - E_{sup}. \tag{16}$$

Eq. (16) implies that reducible supports, which lower the Fermi level, contribute to the adsorption of molecules while oxidizing supports, which raise the Fermi level, improve the desorption of the products. This agrees well with the experimental results on reducible supports [11,14,16].

The analysis of Eq. (9) with taking into account Eqs. (10-16) shows that with lowering the Fermi level at decreasing the size, the reaction rate $R$ increases because the corresponding Fermi level goes down below the antibonding states of oxygen atoms, and their adsorption becomes stronger, leading to the increase of fractions of O atoms. If the size further decreases so the Fermi level goes down below the antibonding states of $CO_2$ molecules, then desorption becomes difficult, and the reaction rate decreases. So, there is an optimal size of the nanoparticles ($r_{opt}$), at which the reaction rate reaches its maximal value. The size of the nanoparticle should be in the range from $r_{CO2}$ to $r_O$ to catalyze effectively the oxidation reaction, corresponding to the Fermi levels equal to antibonding states of $CO_2$ and atomic oxygen, respectively (Fig. 2).

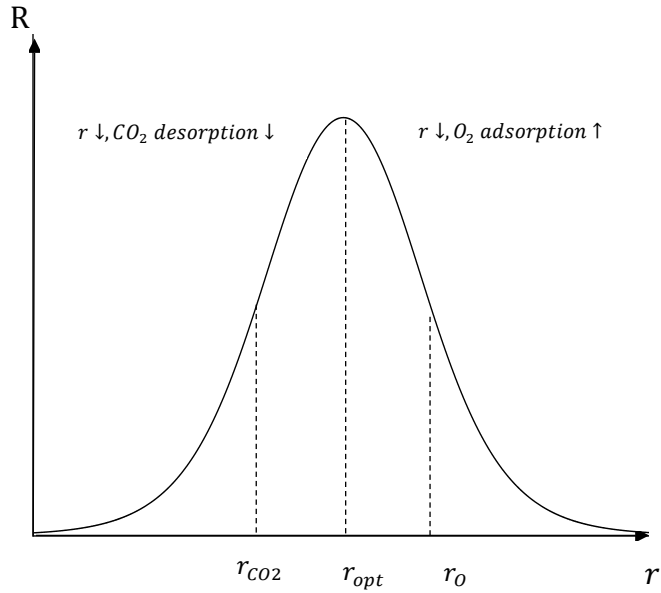

Fig.2. The volcano-shape size dependence of CO oxidation reaction rate for gold nanoclusters

Thus, we derived a volcano-shape dependence for the reaction mechanism of CO oxidation through the initial dissociative adsorption of oxygen molecules, by taking into account the interaction of oxygen molecules with unoccupied active sites (Z) in the kinetic equations at quasi steady-state. Note here, that in the case of $Z \gg N_O$, the dependence of the reaction rate on size is monotonic and does not exhibit a volcano-shape. The only case, when the surface of the catalyst is saturated by atomic oxygen and no dependence of its concentration on the pressure of $O_2$, we can obtain a volcano-shaped plot [22]. Thus, for the reaction rate to exhibit a volcano-shaped size dependence, the interaction of the intermediates with the active sites of the catalyst should be taken into account in the kinetic equations for the intermediates of the reaction.

**Conclusion**

In summary, we obtained a volcano-shaped plot of the dependence of CO oxidation rate on the size of gold nanocatalysts by introducing weak and strong chemisorption states of the intermediates (O, $CO_2$) in the kinetic equations in the framework of the reaction mechanism through the dissociative adsorption of $O_2$ molecules and interaction of atomic oxygen on the surface with CO molecules from the gas-phase. The model is based on the d-band model of the formation of the chemical bond between the adsorbate and the metal surface by taking into account the jellium model for metal clusters. According to the model, upon the initial decrease of particle`s size, the reaction rate increases due to the enhanced adsorption of the reagents and then decreases due to the impaired desorption of the products. The reducible and oxidizing supports can contribute to the reaction rate by lowering or raising the Fermi level of the metal clusters, respectively. We suggest that a volcano-shaped size dependence takes place for the CO oxidation mechanism when interaction of reagents and intermediates with active sites of the catalyst is taken into account in the kinetics of the reaction. Note, that the model presented in the work has been developed for spherical clusters for which Eq. (1) is valid, but we can assume that it can be also applied for the particles of other shapes (for example, ellipsoid or hemisphere particles) with different coefficients in the expression for the ionization potential.